\documentclass[11pt]{article}
\usepackage{graphicx}
\usepackage{placeins}
\usepackage{float}
\usepackage{subfigure}
\usepackage[toc,page]{appendix}

\usepackage{amsmath,amssymb}
\usepackage{epsfig}
\begin{document}
\title{{\bf{\Large Temperature and thermodynamic structure of Einstein's equations for a cosmological black hole}}}
\author{ 
{\bf {\normalsize Krishnakanta Bhattacharya}$
$\thanks{E-mail: krishnakanta@iitg.ernet.in}} \,\ and \,\ 
 {\bf {\normalsize Bibhas Ranjan Majhi}$
$\thanks{E-mail: bibhas.majhi@iitg.ernet.in}}\\ 
{\normalsize Department of Physics, Indian Institute of Technology Guwahati,}
\\{\normalsize Guwahati 781039, Assam, India}
\\[0.3cm]
}

\maketitle

\begin{abstract}
  It is expected that the cosmological black holes are the closest realistic solutions of gravitational theories and they evolve with time. Moreover, the natural way of defining thermodynamic entities for the stationary ones is not applicable in the case of a time dependent spacetime. Here we confine our discussion within the Sultana-Dyer metric which is a cosmological black hole solution of Einstein's gravity. In literature, there exists two expressions of horizon temperature -- one is time dependent and the other does not depend on time. To single out the correct one we find the temperature by studying the Hawking effect in the tunnelling formalism. This leads to time dependent structure. After identifying the correct one, the Einstein's equations are written on the horizon and we show that this leads to the first law of thermodynamics. In this process the expressions for horizon entropy and energy, obtained earlier by explicit calculations, are being used. This provides the evidence that Einstein's equations have thermodynamic structure even for a cosmological black hole spacetime. Moreover, this study further clarifies the correctness of the expressions for the thermodynamic quantities; like temperature, entropy and internal energy.   
\end{abstract}

\section{Introduction and Motivation}
Black hole physics has been one of the most high-yielding domains for the theoretical physicists over the years. Most of the earlier works have been done on the stationary; i.e. time independent black holes in which it is assumed that the spacetime is asymptotically flat. Although those works have been very useful in many cases, those have been suffered from the objection for not caring the realistic situations. In a realistic situation, a black hole should be surrounded by a local mass distribution. Therefore, at large spatial distance from the black hole, the spacetime should not be usually flat. Also, black holes are not usually time independent in a realistic thought. Therefore, one does not treat the stationary ones as the part of a cosmological scenario. Of late, many theoretical efforts have been brought to light in which experts have tried to develop the physics for a dynamical black hole with different levels of success. But, it is yet to be developed in many avenues.

   In this paper we shall discuss about the Sultana-Dyer (SD) black holes \cite{Sultana:2005tp}, metric of which is conformal to the Schwarzschild black hole spacetime and the conformal factor is time dependent. There may be other ways to deform the Schwarzschild metric conformally, but we choose the SD metric due to the fact that it becomes Friedmann-Lemaitre-Robertson-Walker (FLRW) metric in the asymptotic limit, and the FLRW metric describes the homogeneous, isotropic expanding universe very successfully in various cases. Also, it turns out that the SD metric is a inhomogeneous and time dependent solution of general relativity (GR) in presence of two non-interacting perfect fluids. One fluid is a timelike dust and another one is a null dust. Now, as the conformal factor is time-dependent, the spacetime metric evolves with time. For a detailed discussion, see \cite{Sultana:2005tp}. Therefore, one should expect a different result from those calculations for the stationary cases. Recently, the thermodynamical aspects of the metric has been discussed in \cite{Faraoni:2007gq}--\cite{Majhi:2015tpa}. Our motivation in this paper has been to develop the full thermodynamics of the SD black hole in an unprecedented way.

   To develop the thermodynamic relation for a black hole, we need to know the expressions of the thermodynamic quantities (such as temperature, entropy and internal energy) very precisely. The expression of entropy, proposed earlier in \cite{Faraoni:2007gq}, is later on explicitly derived for the SD black hole \cite{Majhi:2014lka,Majhi:2015tpa}. Also, the convenient energy expression, used to describe the thermodynamics of a black hole in evolving space time, is the Misner-Sharp energy \cite{Misner:1964je} . Being spherically symmetric one, the energy for the SD spacetime has been obtained easily in \cite{Faraoni:2014lsa}. Therefore, as far as entropy and energy are concerned, they are uniquely determined. But, there is a discrepancy in the expression of the horizon temperature. In literature we get two expressions of temperature for the SD black hole: one is time independent \cite{Sultana:2005tp} while the other one is time dependent \cite{Saida:2007ru}. In spite of the fact, that the time dependent expression is favoured by the scaling argument \cite{Faraoni:2007gq}, the correct one should be identified by an explicit derivation. One direct way to find the correct expression is to look at the emission spectrum from the horizon. Therefore, it is necessary to study the Hawking effect \cite{Hawking:1974sw} of the black hole. 
   It should be mentioned that the original calculation of Hawking's procedure is
 not applicable here, because, in
the original calculation of Hawking the radiation spectrum was observed by evaluating the Bogoliubov
coefficients of ingoing and outgoing modes. Those coefficients were
defined by the boundary conditions when the spacetime is asymptotically flat. In this case, a time dependent 
conformal factor is added on the Schwarzschild metric which makes
the spacetime to be asymptotically FLRW. Now, it has also been studied that the Bogoliubov coefficients
are not conformally invariant which means that the original method of Hawking is not straightforwardly applicable to study the Hawking radiation for the SD black holes.
   Although one of the authors of the paper obtained the expression of temperature applying the (gravitational) anomaly method and it turned out to be time dependent \cite{Majhi:2014hpa}, it is needed to be verified by some other method. Here, we shall use the tunnelling formalism to derive the expression of temperature of a cosmological black hole horizon from the tunnelling probability. It will be shown that the obtained expression is the same as that, obtained in anomaly method. Before starting our discussions in detail, we would like to make some comments as follows. 

There are two principal ways of implementing the tunnelling formalism, one is the Hamilton-Jacobi method \cite{Srinivasan:1998ty} and the another is the  null geodesic method \cite{Parikh:1999mf}. Both of them are based on the semi- classical WKB approximation and they give the identical result {\footnote{For recent progress and review on tunnelling mechanism, see \cite{Banerjee:2008cf,Banerjee:2009wb,Vanzo:2011wq,Majhi:2011yi}.}}. The inherent idea of tunnelling method is the formation of particle-antiparticle pair near the event horizon. The outgoing positive energy mode is observed as Hawking radiation, whereas the ingoing negative energy mode is trapped inside the horizon. The striking feature in this case is that the SD black holes evolves with time and, therefore, there does not exist any time-like Killing vector to describe the energy of a particle. Rather, we had to introduce the Kodama vector \cite{Kodama:1979vn} to describe this energy while writing the Hamilton-Jacobi equation for the ingoing and the outgoing particles. In our calculations we shall introduce null coordinates in which the ingoing and the outgoing modes are clearly separated out. 

 Next  we shall  find the first law of thermodynamics for the SD black hole. It is now well known fact that the Einstein's equations, written on the horizon, leads to the first law \cite{Kothawala:2007em}. This have been verified for some particular types of black hole spacetimes. Now the question is: Is the same true for the SD metric; i.e. is the Einstein's equations in this case have thermodynamic structure? The answer we shall find here is ``Yes''. For that, we shall project the Einstein's equation on the horizon and observe that it is identical to the first law of thermodynamics.  Interestingly, in this process it will be observed that the already obtained expressions of temperature, entropy and energy are well suited for getting such a relation. This provides a direct evidence of the correctness of the obtained results. 
 
We organise the paper as follows. On the following section (section \ref{SD}) we provide a brief review of the SD metric. Section \ref{Temp} is allotted for the calculation of temperature by the tunnelling approach. After that, in the next section we derive the thermodynamic relation from the Einstein's equations. In the final section we summarise our results and then conclude.

\section{\label{SD}SD metric: a brief review}
     The SD metric is a asymptotically FLRW, cosmological black hole solution of GR in presence two noninteracting perfect fluids: one is timelike and the other one is null-like as sources in the right hand side of Einstein's equations. It turns out that the background metric is conformal by a time dependent conformal factor to the usual static Schwarzschild black hole metric ( for details see \cite{Sultana:2005tp}). The explicit form of the metric is given by \cite{Sultana:2005tp}:
\begin{equation}
ds^2 = a^2(\eta)\Big[-d\eta^2+dr^2+r^2(d\theta^2+\sin^2\theta d\phi^2)+\frac{2M}{r}(d\eta+dr)^2\Big]~.
\label{SD1} 
\end{equation}
The positive constant $M$ is identified as the mass of the Schwarzschild black hole while the conformal factor is given as $a(\eta) = \eta^2$. Here $\eta$, $r$ are the time and the radial coordinates respectively. $\theta$ and $\phi$ are the angular coordinates. The above form of the SD metric can be expressed in Schwarzschild like coordinates as
\begin{equation}
ds^2 =  a^2(t,r)\Big[-F(r)dt^2 + \frac{dr^2}{F(r)}+ r^2(d\theta^2
+\sin^2\theta d\phi^2)\Big]~,
\label{SDSC}
\end{equation}
when one uses the coordinate transformation $\eta = t + 2M \ln(r/2M - 1)$ in (\ref{SD1}). In this case we have $F(r) = 1-2M/r$ and the conformal factor takes the following form \cite{Faraoni:2013aba}:
\begin{equation}
a(t,r) = \Big(t+2M\ln\Big|\frac{r}{2M}-1\Big|\Big)^2~.
\label{a}
\end{equation}
It has been shown earlier \cite{Majhi:2014lka} that the SD metric has a conformal Killing horizon at $r=2M$ where (\ref{a}) diverges.

   The energy-momentum tensor for the source term in Einstein's equations can taken here of the form: $T^{ab} = \mu u^au^b + \tau k^ak^b$. The first term represents the timelike dust with energy density $\mu$ and zero pressure while the last term corresponds to the null source. It has been shown in \cite{Sultana:2005tp} that the energy density of the dust; i.e. $\mu$ is positive when we have $\eta<r(r+2M)/{2M}$. Moreover, in this region the outside observer will see that both the fluids flow radially into the black hole. On the other hand, for late times; i.e. for $\eta>r(r+2M)/{2M}$ the sources become unphysical and the dust becomes superluminal. 
Another issue of ill-behaviour of the sources can be pointed out here in this context. 
 In Ref. \cite{Sultana:2005tp}, it has been shown that the SD black hole is sourced by the dust and the null dust. But such an interpretation  may not be correct as these fluids do not satisfy the conservation laws separately. 
Whatever the case may be, untill another closest realistic solution, got rid of these limitations, is brought to light
one can use SD black hole as a model to explore the realistic cosmological black holes. In that case it is not very important 
whether the matter is ill-behaving in the theory of GR \cite{Faraoni:2014lsa}.
Moreover, despite those limitations one cannot deny that
 the the spacetime metric of SD black hole is one of the closest realistic solution of a cosmological black hole till date and fits nicely to the theory of expanding universe.
  Therefore, it is still very interesting to study in all possible ways to find different aspects. Also, the metric is very simple  and hence we can take it as a model to explore the realistic cases. The unphysical features may be, as cultivated people feel, due to the fact that the solution is far from the realistic one. One can expect if one finds an exact solution that will be free of this problem. However, in the absence of the exact solutions, one needs to take a mode which is close to it. With this spirit we shall consider here the analysis on the SD metric.

\section{\label{Temp}Tunnelling and Hawking temperature}  
    To describe the Hawking radiation by tunnelling mechanism one always takes only the ($r-t$)-sector of the full metric. This is because it has been argued that the tunnelling occurs radially outward from the horizon. In addition to such hand waving argument, there is a much more concrete reason behind this. We know that the Hawking radiation is a near horizon effect. In this region, one already knew that the effective theory reduces to a two dimensional conformal theory and the main physics is driven only by the ($r-t$)-sector of the full metric. As the Hawking radiation is emitted from the horizon, one can take the two dimensional metric for the calculation of the radiation in the tunnelling formalism. 
   
   The near horizon effective theory mostly studied for the static backgrounds. The present metric is time dependent and hence it is not clear if such conclusion can also be drawn for the SD one. Therefore one needs to study the same here. This has already been examined by one of the authors in one of his papers \cite{Majhi:2014hpa}. It has been observed that in this case the near horizon effective metric is given by   
\begin{equation}
 ds^2=a^2(t,r)\Big{[-F(r)dt^2+\frac{dr^2}{F(r)}\Big]}
\label{cal1}
\end{equation}
Now keeping in mind for the future use we want to express the above in null coordinates which are defined as $u=t-r_* $ and $v=t+r_*$ where the tortoise coordinate $r_*$ is given by $dr_*=dr/F$. In these coordinates (\ref{cal1}) turns out to be
\begin{equation}
ds^2=-\frac{a^2(t,r)F(r)}{2}\Big(dudv+dvdu\Big)~.
\label{cal3}
\end{equation}
One of the importances of null coordinates is the ingoing modes and the outgoing modes of the radiation are nicely separated out; which, as we shall observe later on, will be mostly needed in the analysis.

  Now for a general background, the Klein-Gordon equation for massless particle is given by $\nabla_a\nabla^a\phi=(1/\sqrt{-g})\partial_a(\sqrt{-g}g^{ab}\partial_b\phi)=0$. Under the background (\ref{cal3}), this reduces to
\begin{equation}
\partial_u\partial_v\phi=0~.
\label{cal6}
\end{equation}
Hence, the general solution of the equation takes the form $\phi(u,v)=\phi_R(u)+\phi_L(v)$.
The total solution in  null coordinates ($u,v$) is separated out as two different functions, each one is a function of single null coordinate. The subscripts $R$ and $L$  stand for the right and left moving respectively. Later, by calculating the momentum of the corresponding mode, it will be explicitly shown that those separated function corresponds to the outgoing (right) and the ingoing (left) modes of the Hawking radiation at the event horizon. Here lies the importance of null coordinates in the description of Hawking radiation.
As the total wave function is now separated out as the functions of single variable, they should satisfy the relations:
\begin{equation}
\nabla_v\phi_R(u)=0,\,\,\,\ \nabla_u\phi_R(u)\neq 0,\,\,\,\ \nabla_u\phi_L(v)=0,\,\,\,\ \nabla_v\phi_L(v)\neq 0~. 
\label{lr}
\end{equation}
A general way to describe the above relations simultaneously is the equation
\begin{equation}
\nabla_a\phi=\pm\sqrt{-g}\epsilon_{ab}\nabla^b\phi 
\label{cal7}
\end{equation}
which is known as the chirality condition. For more details and usefulness of this condition, see \cite{Banerjee:2008sn}. 
Here the positive sign is assigned for left going mode ($\phi_L(v)$) and the negative sign is assigned for the right moving mode ($\phi_R(u)$). $\epsilon_{ab}$ is the numerical antisymmetric tensor with $\epsilon_{uv}=+1$. 
In ($t,r_*$) coordinates, under the background (\ref{cal3}) the relation (\ref{cal7}) reduces to
\begin{equation}
\partial_t\phi=\mp\partial_{r_*}\phi 
\label{cal8}
\end{equation}
where $\epsilon_{tr}=-1$ has been used.
This is going to be the most important relation throughout our calculations. Here the negative sign is for the left moving mode while the the positive sign corresponds to the right moving mode.

      Next, we need to find the Hamilton-Jacobi equation for these modes. For that one has to first identify the conserved quantity which represents the energy of the particle. For static background, there is a timelike Killing vector $\chi^a$ and the corresponding conserved energy is defined  as $E=-\chi^aP_a$ where $P_a$ is the four-momentum. Since present spacetime is time dependent we can not use this definition. For evolving case the energy of a particle can be defined in terms of the Kodama vector \cite{Kodama:1979vn}:  
\begin{equation}
E=-K^aP_a 
\label{cal9}
\end{equation}
where $K^a$ is the Kodama vector which is defined as \cite{Kodama:1979vn,Vanzo:2011wq}
\begin{equation}
K^i(x)=\frac{\epsilon^{ij}}{\sqrt{-g}}\partial_jR~.
\label{cal10}
\end{equation}
In the above $\epsilon^{ij}$ is the numerical skew tensor with $\epsilon^{tr}=1$ and here $R=ar$. The same has been used earlier in \cite{Cai:2008gw} to study the Hawking radiation from the apparent horizon of FRW universe.
For the metric (\ref{cal3}) the components of the Kodama vector, in ($t,r_*$) coordinates are given by
\begin{equation}
K^t=\frac{1}{a^2F}\partial_{r_*}R; \,\,\,\
K^{r_*}=-\frac{1}{a^2F}\partial_tR~.
\label{cal12}
\end{equation}
It is now well known to us that Hawking radiation is the semi-classical result of quantum field theory in curved space-time. So, we can use the semi-classical WKB ansatz of the wave function i.e. 
\begin{equation}
\phi=e^{\frac{iS}{\hbar}}~,
\label{phi}
\end{equation}
where S is the action. Here, we have intentionally dropped out the normalisation factor of the wave function, which would not play any significant role in the calculation. Therefore, the momentum eigenvalue, in terms of the action, turns out to be $\hat{P}_a\phi = -i\hbar(\partial\phi/\partial x^a) = P_a\phi = (\partial_aS)\phi$ and hence
\begin{equation}
P_a=\partial_aS. 
\label{cal13}
\end{equation}
Earlier, we have stressed on the fact that the Hawking radiation is a near horizon event. Due to pair production, the two modes of radiation are generated. Particle with positive momentum gets out of the surface, whereas, the particle with negative momentum is trapped by the surface. Thus, the outgoing mode and the ingoing mode can be distinguished. This argument we shall use later to show that our earlier sign convention is compatible with the identification of outgoing and ingoing modes. 

    Now, using the explicit expressions (\ref{cal12}) for the components of the Kodama vector in the definition for energy (\ref{cal9}), we obtain
\begin{equation}
E=-\frac{1}{a^2F}(\partial_{r_*}R)(\partial_tS)+\frac{1}{a^2F}(\partial_tR)(\partial_{r_*}S)~. 
\label{cal14}
\end{equation}
On the other hand by the use of the ansatz (\ref{phi}), the chirality condition (\ref{cal8}) reduces to the following form:
\begin{equation}
\partial_tS = \mp\partial_{r^*}S~.
\label{chirality}
\end{equation} 
For the left mode we have $\partial_tS=-\partial_{r_*}S$. Substituting this in (\ref{cal14}) and solving for $\partial_{r^*}S$ we obtain
\begin{equation}
\partial_{r_*}S=\frac{Ea^2F}{\partial_{r_*}R+\partial_tR}~.
\label{cal15}
\end{equation}
Similarly, that for the right moving mode turns out to be
\begin{equation}
\partial_{r_*}S=\frac{Ea^2F}{-\partial_{r_*}R+\partial_tR}~.
\label{cal16}
\end{equation}
Combining them and rewriting in ($t,r$) coordinates we find 
\begin{equation}
\partial_{r}S=\frac{Ea^2}{\pm F \partial_{r}R+\partial_tR}~.
\label{cal18}
\end{equation}
Here the positive sign is assigned for the left mode while the negative sign is for the right mode.
In this case we have $R=ar$ where the expression for $a$ is given by (\ref{a}). With this, the above reduces to the following form:
\begin{equation}
\partial_{r}S=\frac{E}{G(t,r)} 
\label{cal21}
\end{equation}
where
\begin{equation}
G(t,r)=\pm\frac{F}{a}\pm\frac{4M}{a^{\frac{3}{2}}}+\frac{2r}{a^{\frac{3}{2}}}~.
 \label{cal22}
\end{equation}

    Since Hawking radiation is due to the near horizon particle production event, we expand $G(t,r)$ about the horizon $r=r_H$:
\begin{equation}
G(t,r)=G(t,r_H)+(r-r_H)G'(t,r_H)+\dots~. 
\label{cal23}
\end{equation}
In this case $G(t,r_H)$ is calculated from (\ref{cal22}) whereas $G'(t,r_H)$ is evaluated from 
\begin{equation}
G'(t,r)=\pm\frac{F'}{a}+\frac{2}{a^{\frac{3}{2}}}-(\pm)\frac{Fa'}{a^2}-\frac{\frac{3}{2}(2r\pm 4M)a'}{a^{\frac{5}{2}}}~. 
\label{cal24}
\end{equation}
Note that at the horizon $a(t,r)$ diverges while $F(r_H)=0$. So the leading term in the expansion (\ref{cal23}) is due to the first term of (\ref{cal24}). Hence, neglecting all the other terms and keeping only the leading term we obtain
\begin{equation}
G(t,r)\simeq\pm\frac{F'}{a}|_{r_H}(r-r_H)=\pm\frac{2\kappa}{a_H}(r-r_H)~,
\label{G}
\end{equation}
where $\kappa= F'(r_H)/2$ is the surface gravity of the usual Schwarzschild black hole and $a_H$ is the value of the conformal factor at the horizon.
Substitution of this in (\ref{cal21}) yields
\begin{equation}
S=\pm\frac{Ea_H}{2\kappa}\int \frac{1}{r-r_H}dr  
\label{cal26}
\end{equation}
with negative (positive) sign implying the right (left) mode.\\ 
This integration is well defined and real only when the initial and the final point, between which the above integration is performed, are on the same side of the horizon. But, for this case, we want to calculate the transition probability of an outgoing(or ingoing) particle, which is initially at $r<r_H$(or $r>r_H$) and finally reaches $r>r_H$(or $r<r_H$). Therefore we need an extra prescription. But before that, note  that the positive(for ingoing mode) or the negative (for the ingoing mode) signs were determined by the chirality condition and after obtaining equation (\ref{cal26}) we see that it was obvious for physical description. For outgoing particle ( initially at $r<r_H$ ) $\partial_rS>0$, which is only possible when we take the negative sign in the equation (\ref{cal26}).       Same argument is applicable for ingoing particle as well.

  Now, the integration is performed by complex integration method. As in this case the path of the integration actually passes through a singularity $r=r_H$ of the integrand, we must choose a path which avoids the singularity. The evaluation of this integration can be followed from \cite{Book} (See page 60). For outgoing particle we have
\begin{equation}
\int_{r_1}^{r_2}\frac{1}{r-r_H}dr=P\Big(\int_{r_1}^{r_2}\frac{dr}{r-r_H}\Big)- i\pi~,
\label{book1}
\end{equation}
while for ingoing one
\begin{equation}
\int_{r_2}^{r_1}\frac{1}{r-r_H}dr=P\Big(\int_{r_2}^{r_1}\frac{dr}{r-r_H}\Big)+ i\pi~.
\label{book2}
\end{equation}
Here we have chosen the point $r_1$ inside the horizon while $r_2$ is at outside of it.  
The first terms in the above are the principal values of the integrals and these are real. For outgoing (ingoing) mode the contour has been chosen on the upper (lower) half plane. So, for the outgoing mode the action is given by
\begin{equation}
S_{out}=-\frac{Ea_H}{2\kappa}\int_{r_1}^{r_2}\frac{dr}{r-r_H}=\frac{i\pi Ea_H}{2\kappa} + {\textrm{real part}}~. \label{SOUT}
\end{equation}\\
Similarly, for ingoing mode we find
\begin{equation}
S_{in}=\frac{Ea_H}{2\kappa}\int_{r_1}^{r_2}\frac{dr}{r-r_H} = -\frac{Ea_H}{2\kappa}\int_{r_2}^{r_1}\frac{dr}{r-r_H}=-\frac{i\pi Ea_H}{2\kappa} +{\textrm{real part}}~. \label{SIN}
\end{equation}
Substituting them in (\ref{phi}) the probabilities for emission and absorption  are calculated as
\begin{equation}
P_{out}=|\phi_{out}|^2 = |e^{\frac{iS_{out}}{\hbar}}|^2=e^{{- \frac{\pi Ea_H}{\kappa{\hbar}}}} \label{POUT}
\end{equation}
and
\begin{equation}
P_{in}=|\phi_{in}|^2=e^{{ \frac{\pi Ea_H}{\kappa{\hbar}}}}~. \label{PIN}
\end{equation}
Therefore the tunnelling probability turns out to be 
\begin{equation}
\Gamma=\frac{P_{out}}{P_{in}}=e^{{- \frac{2\pi Ea_H}{\kappa{\hbar}}}}~. \label{GAMMA}
\end{equation}
It should be noticed that we have approximated the value of $G(r,t)$ near the event horizon where 
only the leading order term of $G(r,t)$ has been kept. With this approximation the tunnelling rate is similar to the Boltzmann 
factor. Had the value of $G(r,t)$ was not been curtailed upto 
its leading order value, the calculations would end up with the extra terms, contribution of which is substantially little. So, comparing 
to the static case calculation, one can say that in the static case one gets exactly the Boltzmann factor for the calculation 
of tunnelling probability, while, in non-static case only the near horizon approximation leads to the Boltzmann
factor. The similar has also been done earlier (For example, see \cite{Vanzo:2011wq}).
Comparing this with the Boltzmann factor $e^{-\beta E}$ where $\beta$ is the inverse temperature, the temperature of the horizon is identified as
\begin{equation}
T = \frac{\hbar\kappa}{2\pi a_H}~.
\label{temp}
\end{equation}
This expression is identical to what was obtained by anomaly approach \cite{Saida:2007ru,Majhi:2014hpa}. Also this is agrees with scaling argument provided in \cite{Faraoni:2007gq}. 

      Let us now discuss an issue in the context of the tunnelling approach. In literatures one can find some ambiguities and anomalies regarding the study of Hawking radiation by tunnelling methods for the stationary black holes. One of the mostly discussed among these is the coordinate dependence of the tunnelling rate. If one use Schwarzschild coordinate, as we have done, the calculation predicts the value of temperature which is twice the Hawking temperature when one defines the tunnelling rate as the amplitude square of the outgoing wave. To solve this anomaly, several theories were erected \cite{Srinivasan:1998ty, Akhmedova:2008dz, VANZO, NADALINI, KIM1, KIM2}. One of them is mentioned in \cite{Srinivasan:1998ty} where the tunnelling rate is defined as the ratio of probabilities of outgoing and ingoing modes. This prescription we have adopted here in our calculations. The other one to solve this factor two problem is introducing the isotopic coordinate and the proper distance along the radial direction which is mentioned in the references \cite{VANZO} and \cite{NADALINI}. In addition to them, it has been shown that if one takes into account the temporal part, then this ambiguity does not arise \cite{Akhmedova:2008dz, Banerjee:2009wb}. Another fruitful approach we want to mention here is the following.
 For static black hole case, one can apply the Rindler coordinates where this problem of factor two does not arise \cite{KIM2}. The idea is the following. The Hawking radiation is due to the pair production near the horizon and in this region the natural coordinates can be taken as the Rindler ones. Therefore, one might say that the Rindler coordinates are the more physical coordinate while studying the Hawking radiation by the tunnelling formalism of a static black hole. 
 It should also be mentioned that the tunnelling rate in Rindler coordinates  is identical to that for the Schwinger mechanism \cite{KIM1, KIM2}.  A comparative study of these two methods has been discussed in various literatures (for example Refs. \cite{Srinivasan:1998ty, KIM1, KIM2} ).  A crude way of saying is: in the Schwinger mechanism, the virtual particle-anti particle pairs are separated by an electric field while in Hawking radiation these are separated by the event horizon or by the geometry of the black holes. This shows that if one discusses the tunnelling method in the same footing like the Schwinger mechanism, there will not be any discrepancy in the value of the horizon temperature.
 But, there are some limitations of using these coordinates. Rindler coordinates can only be applied for the non-extremal black holes. Also, these coordinates are the static ones and are not defined properly for the dynamic cases. Since our metric is time dependent, it is not clear how to define the Rindler coordinates.  For the above mentioned reasons, we have taken the Schwarzschid coordinates and applied the standard prescription of \cite{Srinivasan:1998ty} to obtain the correct expression of temperature as one can see from (\ref{GAMMA}).

     Before concluding this section, let us make the following comments. Note that here temperature is time dependent as $a_H$ depends on time. In \cite{Sultana:2005tp}, the expression was quite different and is given by the Schwarzschild temperature. The reason is as follows. The required conformal Killing vector $\xi^a$ and the conformal factor $\Omega$ should satisfy $\xi^a\xi_a\rightarrow -1$ and $\Omega\rightarrow1$, respectively at the null infinity. But since in the present case $\xi_a\xi^a=-a^2$ and $\Omega=a$, (see  \cite{Majhi:2014lka,Jacobson:1993pf} for details on finding the conformal Killing vector) both does not satisfy the above requirements. Therefore, calculation of the temperature based on this formalism does not give correct answer.
     Another point one should mention that the expression
of temperature has been obtained by studying the only the spherically symmetric mode. It must be remember that Hawking effect is a near horizon phenomenon. As mentioned earlier, in this region the effective theory reduces to a two dimensional conformal theory for static
 as well as non-static background as SD. 
 It has been shown in \cite{Majhi:2014hpa} that if one starts with the massive Klein-Gordon equation in four dimension, it effectively reduces to an equation which is governed by the ($r-t$) sector of the full metric in the near horizon limit. All angular and mass terms do not contribute in this region. Since Hawking radiation is a near horizon phenomenon, we have just used this information in our calculation to derive the temperature. The idea of the dimensional reduction is like this. If one expands the Klein-Gordon equation under a spherically symmetric metric and transforms it in ``tortoise" coordinate $r_{*}$ and makes the partial wave decomposition, one finds that the effective radial potential, which contains the angular part, and the mass term appear with the metric coefficient. Therefore in the near horizon limit, it dies out and the full equation reduces to similar to ($1+1$) dimensional one. More specifically, the near horizon physics can be described by an infinite collection of two dimensional fields each propagating in spacetime with the metric, given by the ($r-t$) section of the full metric. For more discussions on how the effective theory becomes two dimensional conformal theory near the event horizon, we suggest the papers in \cite{2DIM} , which only deals with this issue. In addition to this, it may be noted that if one calculates the tunnelling rate with the full metric then also the same reduces to similar to two dimensional result in the near horizon limit. Such a discussion has already been demonstrated for static case (see the analysis around Eq. (2.31) in \cite{Srinivasan:1998ty}). Here also the angular and mass terms, like the dimensional reduction technique, appear with the metric coefficient which vanishes at the horizon. 
 So as far as the temperature is concerned, one can focus only on the 
spherically symmetric mode which is exactly solvable. This causes no information loss regarding temperature.
If one is concerned about the radiation spectrum or the gray-body factor (the relative factor between the asymptotic
 radiation spectrum and the spectrum
of black body radiation), then one has to take care of all the angular modes, not to loose
the total information. 
     
\section{\label{Einstein1}First law of thermodynamics from Einstein's equation}
It has been observed that black holes behave like thermodynamic objects and satisfy the thermodynamic relations \cite{Bardeen:1973gs}. In the previous section we have found out the expression of temperature ($T$) of the SD background. Also, the expressions of entropy ($S$) and the energy ($E$) (the Misner-Sharp energy) of the black hole are already obtained by explicit calculations \cite{Majhi:2014hpa}. So, the expressions of the thermodynamic quantities ($E,S,T$) are now known to us for the SD background. Now, in this section we shall find out the first law of thermodynamics. It is well known that the Einstein's equations, projected on the horizon, leads to first law of black hole mechanics \cite{Kothawala:2007em}. Which implies that the near horizon field equations of gravity behave like local thermodynamic equilibrium. In this section we want to follow the same strategy in order to find whether the same conclusion can be drawn for a more realistic and time dependent SD black hole. The steps are identical to the earlier work by Hayward \cite{Hayward:1997jp}. The justification of presenting this discussion lies in the fact that it will give an explicit verification of the correctness of the derived thermodynamic quantities. Additionally, a new reader will find the paper as self sufficient.

  The SD metric satisfies the Einstein's equation $G_{ab}=kT_{ab}$, where $G_{ab}=R_{ab}-(1/2)g_{ab}R$ is the Einstein's tensor and $T_{ab}$ is the energy-momentum tensor corresponding to the matter source. Here $k$ is given by $k=8\pi G$ with $G$ is the Newton's constant. The expression for $T_{ab}$ is given by \cite{Sultana:2005tp}:
\begin{equation}
T_{ab}=\frac{1}{\kappa\Omega}(2g_{ab}\nabla^2\Omega -2\nabla_a\nabla_b\Omega -3\Omega^{-1}g_{ab}g^{mn}\nabla_m\Omega\nabla_n\Omega)~,
 \label{EMT}
\end{equation}
where in our notations, $\Omega=a(t,r)$. Using the above relation, the explicit form of the required components of the energy-momentum tensor for the present black hole (\ref{SDSC}) are given by
\begin{equation}
k T_r^r=\frac{-2\ddot{a}}{a^3F}+\frac{5(\dot{a})^2}{a^4F}+\frac{a'F'}{a^3}-\frac{F(a')^2}{a^4}~, 
\label{TRR}
\end{equation}
and
\begin{equation}
k T_t^t=\frac{2Fa''}{a^3}-\frac{5F(a')^2}{a^4}+\frac{a'F'}{a^3}-\frac{(\dot{a})^2}{a^4F}~.
\label{TTT}
\end{equation}
Next, the explicit form of the relevant components of Einstein's tensor under the same background turn out to be
\begin{equation}
G_r^r=\frac{-1+F+rF'}{r^2a^2} +\frac{3F(a')^2}{a^4}+\frac{(\dot{a})^2}{a^4F}+\frac{4Fa'}{a^3r}+\frac{F'a'}{a^3}-\frac{2\ddot{a}}{a^3F} 
\label{GRR}
\end{equation}
and
\begin{equation}
G_t^t=\frac{-1+F+rF^{'}}{r^2a^2}+\frac{F'a'}{a^3}+\frac{4Fa'}{a^3r}+\frac{2Fa''}{a^3}-\frac{F(a')^2}{a^4}-\frac{3(\dot{a})^2}{a^4F}~. 
\label{GTT}
\end{equation}
Here, we used the notations as $a'=\partial_ra$, $a''=\partial_r^2a$, $\dot{a}=\partial_ta$ and $\ddot{a}=\partial_t^2a $. Using the Einstein's equation $G_r^r=k T_r^r$ one gets 
\begin{equation}
\frac{-1}{r^2a^2}+\frac{F}{r^2a^2}+\frac{F'}{ra^2}+\frac{4Fa'}{a^3r}+\frac{4F(a')^2}{a^4}-\frac{4(\dot{a})^2}{a^4F}=0 
\label{GTT=TTT}
\end{equation}
The same expression is also obtained when one uses another Einstein's equation $G_t^t=k T_t^t $. Now, we want to evaluate the above expression at the horizon $r=2M$. Therefore, one should examine each term and find whether the term contributes at the event horizon $r=2M$. Since the second term is containing $F$ at the numerator, it must vanishes at the horizon. Denominators of the first and the third terms diverges as $a^2$ in this limit. Since $a'$ is given by $a'=({4M\sqrt{a}})/({rF})$, the denominator of the fourth term diverges as $a^{5/2}$. Use of $\dot{a}=2\sqrt{a}$ leads to the fact that the last two terms vanishes as $a^3F$ at the horizon; i.e. they are the diverging. But it is interesting to note that their divergence is in the same order and, since they are opposite in sign, the collective contribution from these two terms is zero in the near horizon limit. Nevertheless, a graph has been plotted below to convince people that those two terms really do not contribute as a whole at the event horizon. 
\begin{figure}[H]
  \centering
    \includegraphics[width=0.5\textwidth]{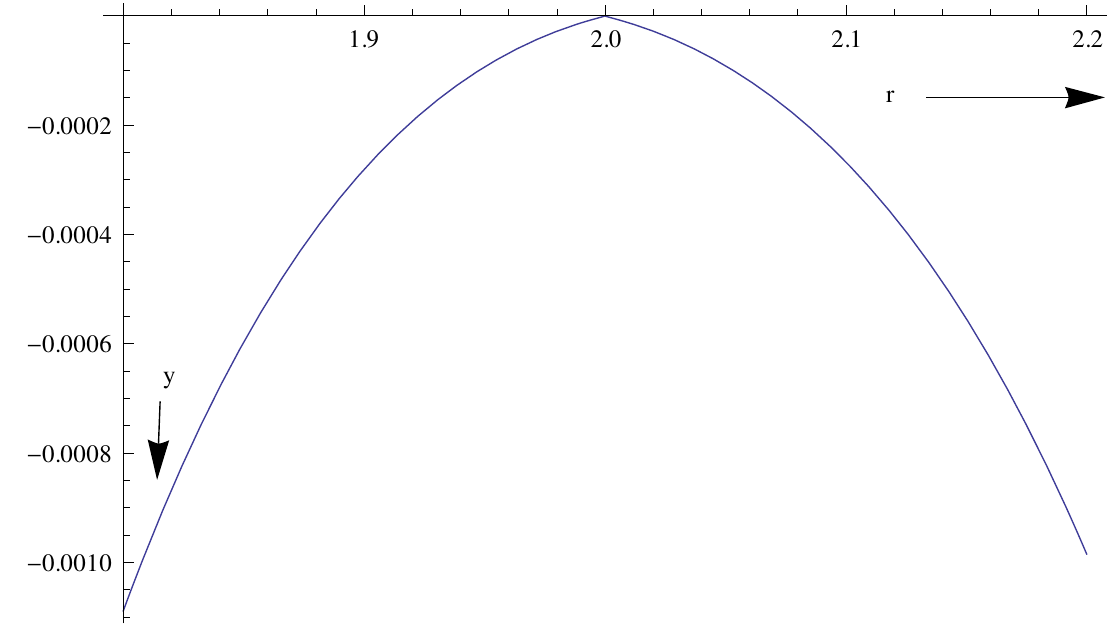}
     \caption{{\it{$y=\frac{4F(a')^2}{a^4}-\frac{4(\dot{a})^2}{a^4F}$ Vs $r$ plot for $t=1=M$.}}}
\end{figure}
\noindent
\linebreak
Therefore, keeping only the dominating terms (the first and the third term which varies with $a^{-2}$ at the horizon) in (\ref{GTT=TTT}) in the near horizon limit, we obtain
\begin{equation}
-1+r_HF'(r_H)=0~.
\label{Einstein}
\end{equation}
Next multiplying each term with $d(a_Hr_H)/2$ and then inserting $\pi$ factor in both numerator and denominator in the second term, one obtains 
\begin{equation}
-\frac{d(a_Hr_H)}{2}+\frac{F'(r_H)}{4\pi a_H} d[\pi (r_H a_H)^2]=0~.
\label{TdS}
\end{equation}
Using the expressions for energy (Misner-Sharp energy) and entropy $E=(r_Ha_H)/{2}$ and $S=\pi r_H^2a_H^2$, respectively, derived earlier in \cite{Majhi:2014hpa} and the expression of temperature, given by (\ref{temp}) we can rewrite the above as
\begin{equation}
dE=TdS~, 
\label{1stlaw}
\end{equation}
which is the first law of black hole mechanics for the SD metric.

   In this section we found that the Einstein's equations, for the SD spacetime, have a thermodynamic structure like the usual cases. The radial-radial component as well as the time-time component, both leads to the same equation at the horizon which ultimately reduces to the first law of mechanics. In doing this it has been observed that the derived temperature (the time dependent one, derived in this paper), and the entropy and energy (found earlier in literature) led to the correct form of thermodynamic law. This once again is the signature of the correctness of these thermodynamic entities.
   
    Let us now mention that the present approach of finding the first law of thermodynamics is completely equivalent to the earlier one by Hayward \cite{Hayward:1997jp}.  
 In the mentioned paper of Hayward, the unified form of the first law of thermodynamics has been given for a spherically symmetric background. In doing so, first he has defined two quantities: energy density and energy flux. Using them and taking into account the Misner-Sharp energy, the required result has been obtained. The main idea is as follows. The gradient of the Misner-Sharp energy, manipulated by using the Einstein's equation, when projected on the horizon leads to the unified first law in the differential form.
So it is obvious that the present analysis is in same line of Hayward. But we present this not only for the completeness of the discussion, but also for a new reader who can find the paper in a self-contained manner. Moreover, as we have mentioned earlier, the expression of temperature for the SD black hole is a long lasting problem in literature, causing confusion while giving thermodynamic description of the black hole. Introducing the tunnelling formalism we have found out the expression of temperature, which turns out to be the time dependent. Now, we had to verify whether the expression was correct. To see it explicitly we presented the above analysis. Here, we had taken the Einstein's equation which we have projected near the event horizon and taking the leading order terms we have shown that the first law of thermodynamics can be achieved for the SD ones. In that process we have shown that the obtained value of the temperature fits nicely to the expression of entropy and the Misner-Sharp energy of the literature to give the first law. 
 The analysis shows that the Einstein's equation manifests itself as a thermodynamic identity when it is projected to the horizon and this work revels the thermodynamic correspondence of Einstein's equation in a more obvious manner. Moreover, we have shown the thermodynamic structure of the SD black hole in a more explicit way with proper thermodynamic variables (energy, entropy and temperature) defined for the SD black hole. Not only this, the obtained thermodynamic expression (which is indeed the first law) has been able to solve the long lasting confusion about the exact expression of temperature and we can claim that our obtained expression of temperature is the correct one which is consistent to the other thermodynamic variables for the SD black hole given in other literature. 
 Moreover, our analysis shows that   
there is a deeper connection between spacetime geometry and the thermodynamics of a black hole 
(or precisely thermodynamics of the horizon). In a broader picture this can have the following implications. 
Originally, the concept of entropy and temperature of the different kinds of horizons
in general relativity were well developed, and physicists remarked that the near horizon
behaviour of the field equations of gravity is like local thermodynamic equilibrium in thermodynamics (For more details, see \cite{Kothawala:2007em}).
But, the people contemplated that the spacetime geometry of a black hole is more fundamental compared to the thermodynamics 
and that connection of geometry with the thermodynamics might not be obtained for the more realistic cases where
the spacetime is more complicated to explore and time dependent.
Whereas, we have shown that in a more realistic model one can get the same connection of geometry and thermodynamics
via gravitational equation. Therefore, one can conclude that the thermodynamic description of a black hole is no more
less fundamental comparing to the geometric description of the black hole and the thought, that the near horizon
behaviour of the field equations of gravity might appear like local thermodynamic equilibrium, is still
applicable for a time dependent black hole as well{\footnote{A time dependent charged solution in the expanding universe has been given in \cite{MAEDA}.}}. Moreover, the present one implies that the gravity can be thought as a long wavelength, emergent phenomenon,
and gravitational description, therefore, resembles
to the equations of thermodynamics .

\section{\label{Summary}Summary and Discussions}
   In this paper we have tried to shed light on the thermodynamic aspects of the time dependent
Sultana-Dayer(SD) black hole and thereby obtaining the informations to describe the
SD black hole thermodynamically. This metric, conformally connected to
the usual Schwarzschild one with a time dependent conformal factor, is
asymptotically FLRW which is widely used to describe homogeneous, isotropic
expanding universe. We have provided a brief review on the SD metric before starting
our discussions in details. In the earlier part of our discussion in this paper, we
have bequeathed ourselves to obtain the exact expression of the temperature. We have already mentioned about the incongruity that one observes in
literatures in the expression of temperature. Therefore, we
had to get rid of this conflicting situation and verify which expression is the
correct one for this black hole, the expression in which the temperature depends
on time or the expression in which it is independent of time. For that, we have
studied the Hawking radiation by tunnelling formalism and after overcoming all the
hurdles in calculations we have proved that the expression of temperature of the SD
black hole really depends on the time. Moreover, the expression of
temperature that we obtained from the tunnelling formalism resembles to the same,
obtained by gravitational anomaly approach \cite{Majhi:2014hpa}. Also, this expression of temperature
accords with the scaling argument \cite{Faraoni:2007gq} as we have mentioned earlier.

In the later part of this paper, we were prone to realise whether one can procure
any information from the Einstein's equation to describe the SD black hole
thermodynamically. Our inspiration was the fact that the Einstein's equation leads
to the first law of thermodynamics when it is projected to the horizon, though
verified only for some particular black holes. Here, in this paper we have proved
that similar expression is obtained for the time dependent SD black hole as well. We
have also pointed out the fact that the already obtained expressions of entropy and energy nicely fits with our obtained value of temperature to give the first law
of thermodynamics from the Einstein's equation.

To summarize, we  can draw two major conclusions from our whole discussion. Firstly,
the expression of temperature depends on time and it is consistent to all other
obtained expressions of the thermodynamic quantities such as entropy, energy etc.
Secondly, thermodynamic description is embedded in the Einstein's equation for
the time dependent SD black hole as well, implying that the thermodynamic description and the geometrical description of a black hole should be treated in equal footing.

\vskip 9mm
\section*{Acknowledgments}
The research of one of the authors (BRM) is supported by a START-UP RESEARCH GRANT (No. SG/PHY/P/BRM/01) from Indian Institute of Technology Guwahati, India.

\end{document}